# ON THE RADIATION FROM UNBAFFLED PISTONS AND THEIR DIPOLE EQUIVALENT

*Filipe Soares[1*], Vincent Debut[2,3]*

[1] Instituto Superior Técnico, Centro de Ciências e Tecnologias Nucleares, Lisboa, Portugal
[2] Instituto de Etnomusicologia, Centro de Estudos em Música e Dança, Universidade Nova de Lisboa, Lisboa, Portugal
[3] Instituto Politécnico de Castelo Branco - Escola Superior de Artes Aplicadas, Castelo Branco, Portugal

**ABSTRACT**

The radiation efficiency from simple vibrating planar surfaces is often used as a basis to describe the sound radiation from more complex structures, having important applications in various fields of acoustics. The low-frequency radiation efficiency of a baffled piston can easily be represented by a simple monopole source. Notably, the equivalent source strength is dependent on the piston surface area. However, the unbaffled case presents additional difficulties as the so-called "edge effects" significantly alter the piston radiation impedance. Consequently, a low-frequency equivalence between dipoles and an unbaffled pistons is not as straight forward, since not only the piston area but also its shape will have an effect on the radiated sound. In this work, the search for a simple and generic, equivalence between dipoles and unbaffled pistons is pursued. A finite element model is used to calculate the radiation efficiency from unbaffled pistons with the same surface area but different shapes. A broad set of results indicate that the "edge effects" can be accurately represented by a simple term dependent on the piston compactness (ratio of area to perimeter). Effectively, pistons with smaller area to perimeter ratio will be less efficient radiators. Such term allows the definition of an equivalent dipole source strength that approximates the low-frequency behavior of an unbaffled piston of arbitrary shape.

## 1. INTRODUCTION

Unbaffled pistons of any shape, vibrating at a frequency $\omega$, will radiate approximately as dipoles when the wavelength $\lambda$ is much larger than a characteristic dimension of the piston $a$ [1]. The same analogy can be made with respect to baffled pistons and monopoles. At low-frequencies, the equivalent monopole source strength representing a baffled piston is dependent on the piston area $S$. However, the overall shape of the piston has no influence on the amount of radiated power. Hence, representing the low-frequency radiative behavior of a baffled piston (of any shape) using a monopole becomes a fairly straight forward task.

Attempting to define an equivalence between a dipole and an unbaffled piston is, however, not as trivial. Difficulties arise because the same independence on the piston shape does not hold for the case of an unbaffled piston. That is, pistons with the same surface area $S$ but different shapes will not radiate the same amount of acoustic power. This is related to the effects along the piston edges, which present large tangential velocities and significantly influence the force applied on the fluid by the vibrating piston [2]. In this study we explore how the shape of an unbaffled piston can influence its radiation efficiency.

### 1.1 The monopole

A monopole source, with harmonically varying source strength $q(t) = \text{Re}(Qe^{i\omega t})$, generates a pressure field given by [1]

$$p(r,t) = \frac{Q}{4\pi r} e^{i(\omega t - kr)} \quad (1)$$

where $k = \omega/c$, $c$ is the speed of sound and $r$ is the distance between the monopole and the observation point. The total radiated sound power is given by

$$\Pi = \left(\frac{1}{\rho c}\right) \frac{|Q|^2}{8\pi} \quad (2)$$

where $\rho$ is the fluid density. If we consider a baffled piston of area $S$, vibrating vertically with velocity $u(t) = Ue^{i\omega t}$, the equivalent monopole strength is given simply by

$$Q = i\omega\rho S U \qquad (3)$$

The radiated power of such piston is then given by

$$\Pi = 2\left(\frac{1}{\rho c}\right)\frac{|Q|^2}{8\pi} = \frac{k^2 \rho c S^2 U^2}{4\pi} \qquad (4)$$

where a factor of 2 is considered to account for the effect of the baffle. At low frequencies, its radiation efficiency (normalized radiated power) is given by [1]

$$\sigma = \frac{\Pi}{\rho c S U^2} = \frac{k^2 S}{4\pi} \qquad (5)$$

A baffled piston at low-frequencies, like a monopole source, radiates sound by forcing an unsteady volume outflow from a region that is very small compared to the wavelength. Any vibrating body undergoing a change in volume (e.g. pulsating bubble, baffled loudspeaker, etc.) falls within this category. Consequently, the shape of a baffled piston does not influence its total radiated power, and its radiation efficiency is dependent solely on the surface area $S$ and frequency $k$.

**1.2 The dipole**

Similarly, a dipole radiating harmonically generates a pressure field given by [1]

$$p(r,\theta,t) = \frac{ik}{4\pi r}\cos\theta\left(1+\frac{1}{ikr}\right)Fe^{i(\omega t - kr)} \qquad (6)$$

where $\theta$ is the elevation angle and the source strength here is $f(t) = Fe^{i\omega t}$ and represents a point force. Unlike the cases of the monopole and baffled pistons, a dipole produces no net volume outflow. However, it exerts a force $f(t)$ on the surrounding fluid. A rigid sphere or planar surface oscillating in free space will behave as a dipole when their characteristic dimensions are small compared to the wavelength. The total sound power radiated by this dipole is given by

$$\Pi = \left(\frac{1}{\rho c}\right)\frac{k^2|F|^2}{24\pi} \qquad (7)$$

*1.2.1 Dipole as a superposition of monopoles*

A dipole can also be described by a pair of nearby monopoles radiating in phase opposition, separated by a small distance $d$. In this analogy, the pair of monopoles is equivalent to a dipole source provided that $d \ll \lambda$ and the following equivalence between the dipole and monopole strengths can be made

$$f(t) = Qde^{i\omega t} \quad \Leftrightarrow \quad F = Qd \qquad (8)$$

Similar to the above analogy between monopoles and baffled pistons, an unbaffled piston will radiate as a dipole. The radiated power is given by

$$\Pi = \left(\frac{1}{\rho c}\right)\frac{k^2|Qd|^2}{24\pi} = \left(\frac{k^4 \rho c S^2 U^2}{24\pi}\right)d^2 \qquad (9)$$

while the radiation efficiency is

$$\sigma_{\text{dip}} = \left(\frac{k^4 S}{24\pi}\right)d^2 \qquad (10)$$

Notice how radiation efficiency of the unbaffled case depends on $k^4$ compared to the baffled case $k^2$, i.e. the presence of the baffle will increase the radiation efficiency of a vibrating piston, particularly at low frequencies. More importantly however, unlike the case of the baffled piston, the shape of the piston will have an important effect on its radiation efficiency. In the analogy to the pair of monopoles, the "edge effects" are represented here by the distance $d$, whose physical significance when dealing with real systems (with characteristic dimensions) is not evident. The equivalence of this distance $d$ for the case of physical systems, is not straight forward, particularly since very few analytical examples provide a basis for comparison. Nevertheless, a known example is that of a vibrating sphere [1] of radius $a$ which, at low frequencies, radiates as a dipole whose equivalent distance is $d = 2a$. Another example is that of a rigid disk oscillating in free-space [3], which gives an equivalent distance $d = (8/3\pi)a$. Explicit expressions for these two examples are given in the appendix.

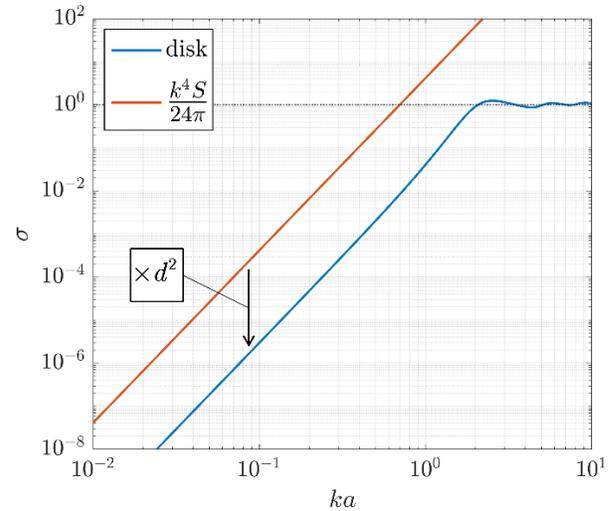

**Figure 1.** Radiation efficiency by a rigid disk of radius $a$ oscillating in free-space (blue) and a low-frequency approximation using a dipole source equivalence (red). The role of the parameter $d$ is highlighted.

The main goal of this study is to provide a simple way to describe the radiation efficiency of an unbaffled piston of arbitrary shape using a point dipole. This aim can be summarised as finding an equivalent distance $d$, dependent on the piston geometry, that allows a reasonable approximation of the radiation efficiency at low-frequencies. To graphically illustrate the effect of the distance $d$, Figure 1 shows the radiation efficiency of a disk in free-space alongside the low-frequency approximation based on the dipole expression (10).

To this aim, a finite-element model was used to calculate the radiation efficiency of unbaffled pistons of various shapes. Based on the obtained results, we derive a generic expression for the equivalent distance $d$, dependent on the compactness (ratio of area to perimeter) of the planar surface, which provides accurate approximations for the radiation efficiency of unbaffled pistons of arbitrary shape.

## 2. NUMERICAL MODEL

The conducted numerical experiments aimed at calculating the radiated power and radiation efficiency of pistons with the same surface area $S$ but different perimeters, in an attempt to characterise/quantify the "edge effects". Three distinct shapes, illustrated in Figure 2, were considered: a rectangle, an isosceles triangle and an ellipse. In this way, similar shapes with different ratios of area to perimeter were calculated. Note that in all three cases, a maximum in the area/perimeter ratio is reached when the aspect ratio equals one $(a/b = 1)$, leading to a square, an equilateral triangle and a circle, respectively.

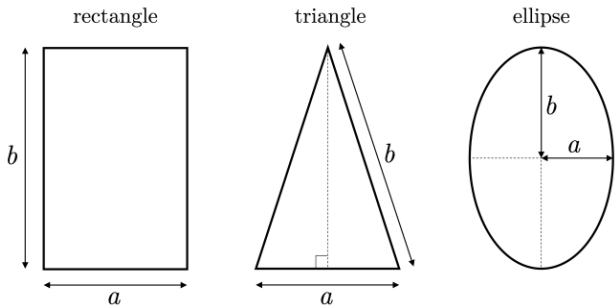

**Figure 2.** Three piston shapes considered in this study.

The used finite element model is illustrated in Figure 3. It consisted of a spherical domain with the unbaffled piston located at the centre of the sphere. A perfectly matched layer is placed in the outer surface of the sphere to simulate anechoic conditions. Note that the spherical domain can be reduced to a quarter of a hemisphere (one eighth of the sphere) by considering the symmetric properties of the problem. Firstly, by the nature of the unbaffled problem, the plane in which the piston is placed will have zero pressure (sound waves in each side cancel out), hence only the upper hemisphere is considered (the pressure in the lower hemisphere will be the same, with opposite sign). Furthermore, the rectangle and ellipse have a two symmetry lines, hence, symmetric boundary conditions are imposed and only one quarter of the hemisphere needs to be considered. For the triangle, which has only one symmetry line, half the hemisphere needs to be considered. The model solve the Helmholtz equation and relies on forcing the piston surface to oscillate vertically at a given velocity (taken as $U = 1 \text{ m/s}$) and frequency $k$, and then calculating the total radiated sound power $\Pi$ via integration over the entire spherical outer surface. Subsequently, the radiation efficiency is given simply by $\sigma = \Pi / \rho c S U^2$.

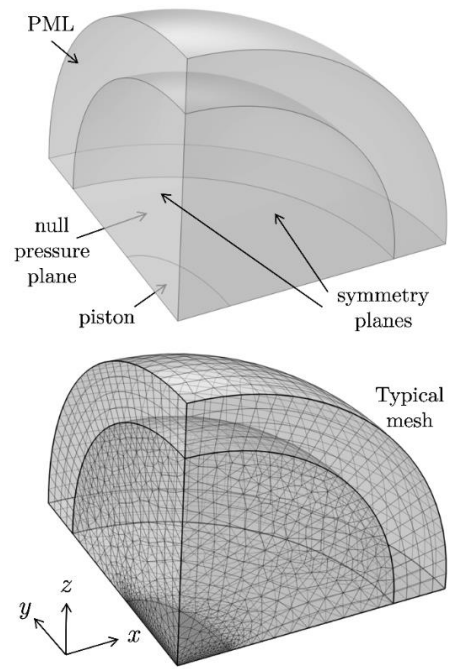

**Figure 3.** Illustration of the used FE model.

## 3. RESULTS: RADIATION EFFICIENCY OF UNBAFFLED PISTONS OF VARIOUS SHAPES

In the search for an equivalent dipole formulation of an unbaffled piston we are naturally interested in the low-frequency behaviour only, i.e. $ka < 1$. The dipole point source cannot, by default, reproduce the radiative behaviour when the wavelength is comparable with the dimensions of the piston (the dipole does not have a characteristic dimension). We proceeded with numerical simulations for all piston shapes (with the same surface area $S$) at a single low-frequency, fixed at $kl = 0.01$, where $l$ was defined as $l = \sqrt{S/\pi}$ (Note: this definition is equivalent to say $l = a$ for the case of a circular piston). Figure 4 shows the radiation efficiency at $kl = 0.01$ for the three base shapes with different aspect ratios $a/b$.

The results in Figure 4 illustrate how pistons with the same surface area and different shapes can have very different radiative capacity. Firstly, we note that the radiation efficiency tends to zero at the limiting cases: $a/b \to 0$, $a/b \to \infty$ and $a/b \to 2$ for the triangle. This seems physically plausible since an infinite "string" will radiate no sound. Secondly, for all three cases, the maximum in radiation efficiency occurs at $a/b = 1$. Note that it is at this point $(a/b = 1)$ that the perimeter of each shape is minimum.

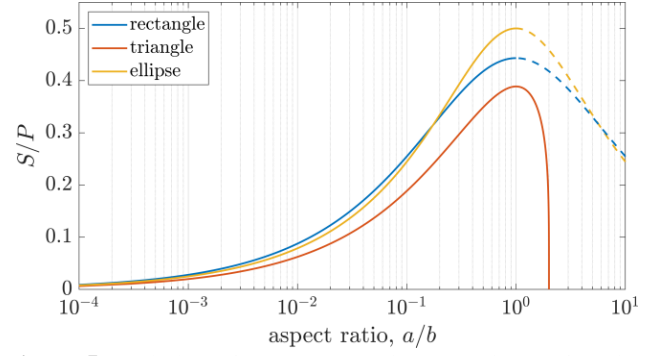

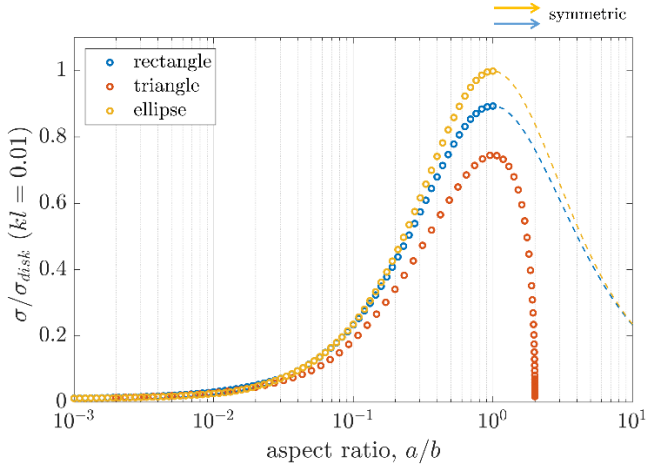

**Figure 4.** Radiation efficiency at low-frequency $kl = 0.01$ as a function of the aspect ratio for the three base shapes (rectangle, triangle and ellipse). Note: the radiation efficiency was normalized by that of a rigid disk in free-space (see appendix A.1).

Another important observation is that all shapes have different values for the maximum of $\sigma$. Notably, the circle is the most efficient, followed by the square and lastly the equilateral triangle. These results strongly suggest that the "compactness" (ratio of area to perimeter $S/P$) is an important parameter in defining the radiation efficiency. This is illustrated clearly in Figure 5, where the compactness parameter $S/P$ of all shapes is plotted as a function of the aspect ratio. Perhaps not so surprisingly, this relation is almost identical to the plot shown in Figure 4. Notice also how this parameter also has units of meters, much like the equivalent distance $d$. Analytical expressions for the areas, perimeters and their ratio for all three shapes are presented in the appendix.

**Figure 5.** Variation of the area to perimeter ratio as a function of the aspect ratio for all three shapes. In these results, the surface area was fixed for all cases at $S = \pi$.

## 4. A GENERIC APPROXIMATION OF THE EQUIVALENT DISTANCE

We now present the same results in a different perspective. Firstly, instead of underlining the role of the aspect ratio $a/b$, we plot the results in terms of the compactness (ratio of area to perimeter $S/P$). Notice however that the ratio $S/P$ is not independent of scale and, in our simulations, both dimensions $a$ and $b$ were modified to ensure a constant surface area $S$. Then, a more pertinent parameter is the normalized compactness $S/Pb$, which is dimensionless and depends solely on the aspect ratio and the type of shape (see the expressions in the appendix, which highlight the appearance of this dimensionless parameter). Secondly, we normalize the radiation efficiency results to that of the dipole $s_0 = k^4 S / 24\pi$, to emphasize how the equivalent distance $d$ can be retrieved, that is

$$\sigma_{FE} = \sigma_{\text{dip}} = s_0 d^2 = \left(\frac{k^4 S}{24\pi}\right) d^2 \Leftrightarrow d = \sqrt{\frac{\sigma_{FE}}{s_0}} \quad (11)$$

Once again, the distance $d$ is not independent of scale and a more generic representation is given when normalized by a characteristic dimension of the piston $d^* = d/b$.

Figure 6 shows the influence of the piston compactness on its radiation efficiency (represented here by the dimensionless equivalent distance $d^*$. Results in the two plots are the same: the top-plot has linear y-axis while the bottom-plot has logarithmic y-axis. From Figure 6 there is no doubt that the compactness of the piston is directly related to the equivalent distance $d$. Considering this parameter, the numerical results for all three cases collapse almost perfectly. Taking the pivot point (known result for the disk, $d^* = 8/3\pi$) we can easily fit the numerical results by

$$d = 2\left(\frac{8}{3\pi}\right)\left(\frac{S}{P}\right) \approx 1.7\left(\frac{S}{P}\right) \quad (12)$$

This curve is shown in Figure 6 as a dotted black line. Note that this expression is exact for the case of the disk and fits almost perfectly to the numerical results of the ellipse (error <2% for aspect ratios $a^* > 0.1$). Slight deviations are seen compared to the rectangle and triangle, but these are no larger than $< 6\%$ for reasonable aspect ratios $0.1 < a^* < 10$.

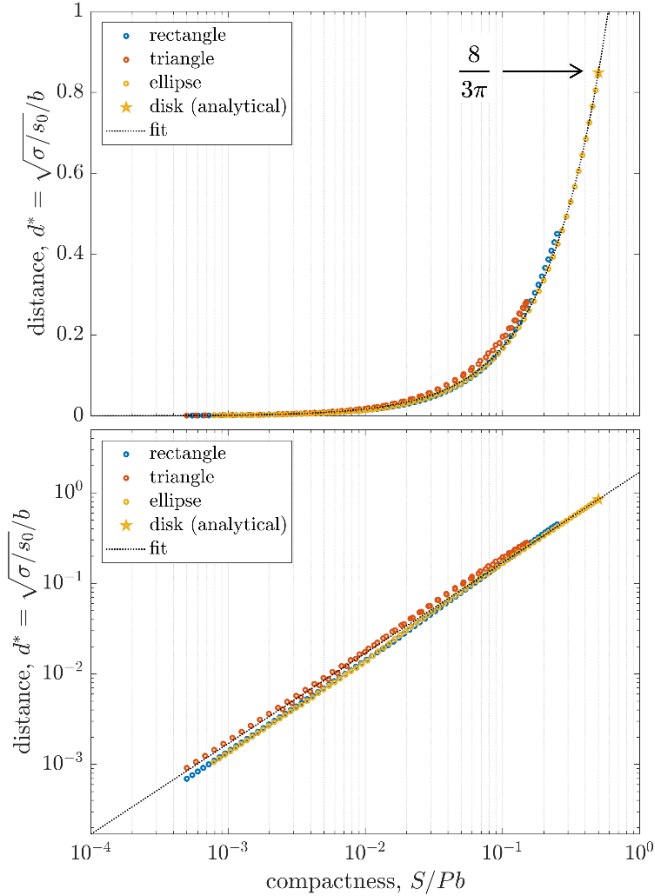

**Figure 6.** Influence of the piston (dimensionless) compactness $S/Pb$ on the radiation efficiency (here represented by the equivalent dimensionless distance $d^* = d/b$). The two plots show the same results with either a linear (top) and logarithmic (bottom) scale for the y-axis. Results for all three shapes collapse in terms of the compactness parameter $S/Pb$.

## 5. CONCLUSIONS

The radiation efficiency of an unbaffled piston with arbitrary shape was studied. Focus was given on providing a simple and generic formulation allowing the description of its low-frequency behavior using a point dipole. A 3D finite-element model was used to calculate the radiation efficiency of various pistons with simple shapes (ellipse, rectangle and triangle) at different aspect ratios. Numerical results strongly suggest that the piston compactness parameter (area to perimeter ratio) is an excellent descriptor of the typical "edge effects" of unbaffled configurations, as numerical results for all shapes collapse as a function of this parameter. Consequently, an expression for the source strength of an equivalent dipole, dependent on the compactness parameter, was provided. This allows for an accurate description of the low-frequency behavior of an unbaffled piston of arbitrary shape, using a simple dipole formulation. The results presented here can be useful in the description of more complex unbaffled structures (e.g. loudspeakers, vibrating plates, radiating elements in musical instruments, etc.), that are often difficult to calculate [4], in a very simple manner. Notably, this could serve as a basis for formulations relying on the use of a minimal number of elementary sources to describe the radiation from arbitrarily shaped plates, as done recently for the baffled case using circular pistons [5].

## APPENDICES

### A.1 Radiation of an oscillating rigid disk in free-space

At low frequencies, the radiation impedance of a rigid disk in free-space is given by [3]

$$Z_R = 2\rho c S \left( \frac{8}{27\pi^2}(ka)^4 + i\frac{4}{3\pi}(ka) \right) \quad (13)$$

and the radiated power is retrieved by

$$\Pi = \frac{1}{2}\text{Re}(Z)U^2 = (ka)^4 \rho c S \left( \frac{8}{27\pi^2} \right) U^2 \quad (14)$$

Re-formulating this expression to emphasize its equivalence to the dipole leads to

$$\Pi = \frac{k^2}{\rho c} \frac{|Qd|^2}{24\pi} = \frac{k^2}{\rho c} \frac{(\omega \rho SU)^2}{24\pi} \left(\frac{192}{27\pi^2}\right) a^2 \quad (15)$$

Then the equivalent distance $d$ becomes evident

$$d^2 = \frac{192}{27\pi^2} a^2 \Leftrightarrow d = \frac{8}{3\pi} a \quad (16)$$

The curious reader might note that the factor $8/3\pi$ is rather familiar. It appears also in the low-frequency reactance of a baffled circular piston [2]. The low-frequency radiation efficiency of a free-disk oscillating in free space is then given by

$$\sigma_{disk} = \frac{(ka)^4}{24} \left(\frac{8}{3\pi}\right)^2 \quad ka < 1 \quad (17)$$

**A.2 Radiation of an oscillating rigid sphere in free-space**

The impedance of a rigid sphere vibrating in free-space is given by [1]

$$Z_R = \rho c \frac{2\pi a^2}{3} \left(\frac{ika(1+ika)}{1+ika-(ka)^2/2}\right) \quad (18)$$

and the radiated power is retrieved by

$$\Pi = \frac{1}{2}\text{Re}(Z)U^2 = \frac{\pi}{6}\rho c a^2 U^2 \left(\frac{(ka)^4}{1+(ka)^4/4}\right) \quad (19)$$

with the associated low-frequency approximation

$$\Pi = \frac{\pi}{6}\rho c a^2 U^2 (ka)^4 \quad ka < 1 \quad (20)$$

In this case, the equivalent dipole would have a distance $d = 2a$.

**A.3 Areas, perimeters and compactness of simple shapes**

For a rectangle, the area, perimeter and their respective ratio are given by

$$\begin{aligned} S &= ab \ ; \ P = 2(a+b) \\ \frac{S}{P} &= \frac{b}{2}\left(\frac{a^*}{a^*+1}\right) \end{aligned} \quad (21)$$

where $a^* = a/b$. Similarly, for an isosceles triangle we have

$$\begin{aligned} S &= \frac{a}{2}\sqrt{b^2 - \frac{a^2}{4}} \ ; \ P = 2b + a \\ \frac{S}{P} &= \frac{b}{4}\left(\frac{a^*\sqrt{1-a^{*2}/4}}{(1+a^*/2)}\right) \end{aligned} \quad (22)$$

The perimeter of an ellipse has no exact formula. However, a good approximation is given by Ramanujan, leading to

$$\begin{aligned} S &= \pi ab \ ; \\ P &= \pi\left[3(a+b) - \sqrt{(3a+b)(a+3b)}\right] \ ; \\ \frac{S}{P} &= b\left(\frac{a^*}{3(a^*+1) - \sqrt{3a^{*2}+10a^*+3}}\right) \end{aligned} \quad (23)$$